\documentclass[aip,jcp,twocolumn,reprint]{revtex4-1}

\usepackage{graphics,graphicx,amsfonts,amsmath,amsbsy,amssymb,color}
\usepackage{bm}
\usepackage[a4paper,vmargin={20mm,20mm},hmargin={20mm,10mm}]{geometry}
\usepackage{subfigure}
\usepackage{paralist}
\usepackage{bbold}
\usepackage{algorithm}
\usepackage{algorithmic}

\newcommand{\bra}{\langle}
\newcommand{\ket}{\rangle}
\newcommand{\bs}[1]{\boldsymbol{#1}}

\def \Pgen {{P_{\textrm{gen}}}}
\def \Nw {{N_{\textrm{w}}}}
\def \Nspawn {{N_{\textrm{spawn}}}}
\def \Niter {{N_{\textrm{iterations}}}}
\def \Eh {{E_{\textrm{h}}}}
\def \mEh {{\textrm{m}E_{\textrm{h}}}}
\def \Eref {{E_{\textrm{ref}}}}
\def \Evar {{E_{\textrm{var}}}}
\def \EvarPT {{E_{\textrm{var+PT2}}}}
\def \Ep {{E_{\textrm{var+PT2}}^{\textrm{new}}}}

\newcommand{\appropto}{\mathrel{\vcenter{
  \offinterlineskip\halign{\hfil$##$\cr
    \propto\cr\noalign{\kern2pt}\sim\cr\noalign{\kern-2pt}}}}}

\begin{document}

\title{Preconditioning and perturbative estimators in full configuration interaction quantum Monte Carlo}
\author{Nick~S.~Blunt}
\email{nicksblunt@gmail.com}
\affiliation{Department of Chemistry, Lensfield Road, Cambridge, CB2 1EW, United Kingdom}
\author{Alex~J.~W.~Thom}
\affiliation{Department of Chemistry, Lensfield Road, Cambridge, CB2 1EW, United Kingdom}
\author{Charles~J.~C.~Scott}
\affiliation{Department of Chemistry, Lensfield Road, Cambridge, CB2 1EW, United Kingdom}

\begin{abstract}

We propose the use of preconditioning in FCIQMC which, in combination with perturbative estimators, greatly increases the efficiency of the algorithm. The use of preconditioning allows a time step close to unity to be used (without time-step errors), provided that multiple spawning attempts are made per walker. We show that this approach substantially reduces statistical noise on perturbative corrections to initiator error, which improve the accuracy of FCIQMC but which can suffer from significant noise in the original scheme. Therefore, the use of preconditioning and perturbatively-corrected estimators in combination leads to a significantly more efficient algorithm. In addition, a simpler approach to sampling variational and perturbative estimators in FCIQMC is presented, which also allows the variance of the energy to be calculated. These developments are investigated and applied to benzene $(30\textrm{e},108\textrm{o})$, an example where accurate treatment is not possible with the original method.
\end{abstract}

\date{\today}

\maketitle

\section{Introduction}

An important goal of quantum chemistry is the more accurate and routine treatment of strongly correlated systems. For weakly correlated systems, low-order coupled cluster (CC) theory is well motivated and extremely successful\cite{ccsd_t, Bartlett2007}, now pushing to larger systems through adaptations to reduce the scaling of the method\cite{Riplinger2013, Piecuch2009, Eriksen2015}. An ultimate goal is a similarly successful polynomial scaling method for strong correlation, and much work continues in this direction. There is a crucial need for better benchmarks to aid such development.

For this task, methods such as the density matrix renormalization group (DMRG) algorithm\cite{White1992, Chan2004, Olivares2015}, selected configuration interaction (SCI)\cite{Huron1973, Schriber2016, Tubman2016, Holmes2016_2, Garniron2017, Loos2018} and full configuration interaction quantum Monte Carlo (FCIQMC)\cite{Booth2009, Cleland2010, Spencer2012} are important tools. Although they are relatively expensive compared to low-order CC, they are systematically improvable, and capable of providing near-exact benchmarks in regimes where other methods give unsatisfactory results. They are also useful beyond providing benchmarks, for example as complete active space (CAS) solvers in CASPT2 (CAS plus second-order perturbation theory) approaches\cite{Dominika2008, Debashree2008, Thomas2015_3, Manni2016, Smith2017}, or in the case of DMRG, as the method of choice in 1D or quasi-1D systems. Approaches based on coupled cluster theory by including high-order clusters also show promise for this task\cite{Xu2018}.

The current FCIQMC algorithm is time limited far more than it is memory limited. On a large-scale cluster, a large FCIQMC simulation may take multiple days to run, yet use a small fraction of the memory available. Moreover, we usually encounter the situation where the final statistical error is multiple orders of magnitude smaller than systematic error [for example, see Table II of Ref.~(\onlinecite{Blunt2015_3})]. This suggests that it may be possible to devise a faster FCIQMC algorithm in exchange for larger statistical noise, which would be a very desirable trade-off, and there are good reasons to believe that FCIQMC can be made substantially faster than the current algorithm.

We recently demonstrated that it is possible to calculate a second-order perturbative correction to initiator error in FCIQMC\cite{Blunt2018}. This correction can often remove over $85\%$ of initiator error in weakly correlated systems, and can be accumulated from existing information in FCIQMC, and therefore has little extra cost. However for large systems or small walker populations, we find that the associated statistical noise can be very large (the opposite situation to the noise on traditional FCIQMC estimators, described above, where statistical noise is small).

Here we propose a modified algorithm where this situation is greatly improved. Specifically, it is shown that FCIQMC can be performed with preconditioning, as commonly performed in quantum chemistry (and optimization problems generally), which allows the use of a much larger time step. This is achieved at the expense of performing multiple spawning attempts per walker, which limits the savings in computer time overall. However, this regime is highly beneficial for the calculation of the perturbative corrections, often reducing statistical noise by an order of magnitude or more, resulting in a far more efficient algorithm. As such, we show that preconditioning has limited benefits to convergence time in FCIQMC, but significantly helps the calculations of perturbatively-corrected estimators.

In addition, we introduce a more simple and efficient approach to sampling the variational energy, and also demonstrate that it is possible to sample the variance of the energy in FCIQMC, as commonly performed in variational Monte Carlo.

We recap FCIQMC in Section~\ref{sec:fciqmc}. The use of preconditioning in FCIQMC is introduced in Section~\ref{sec:precond_fciqmc}, and then contrasted with the traditional approach in Section~\ref{sec:comparison}. A new approach to calculating estimators is discussed in Section~\ref{sec:estimators_theory}. Lastly, results are given in Section~\ref{sec:results}, investigating perturbative estimators and the preconditioned FCIQMC approach, with an application to benzene.

\section{FCIQMC}
\label{sec:fciqmc}

In FCIQMC the ground-state wave function is converged upon by performing imaginary-time evolution\cite{Booth2009}, where the wave function $| \Psi(\tau) \ket$ obeys
\begin{equation}
| \Psi (\tau + \Delta \tau) \ket = | \Psi (\tau) \ket - \Delta \tau (\hat{H} - E_S\mathbb{1}) | \Psi(\tau) \ket,
\end{equation}
where $\hat{H}$ is the Hamiltonian, $\tau$ denotes imaginary-time and $E_S$ is a shift which is slowly varied to control the walker population. This evolution is performed in a basis, $\{ |D_i\ket \}$, in which the components of $| \Psi (\tau) \ket = \sum_i C_i (\tau) |D_i \ket$ obey
\begin{equation}
C_i(\tau + \Delta \tau) = C_i(\tau) - \Delta \tau \sum_j ( H_{ij} - E_S \delta_{ij} ) C_j(\tau).
\end{equation}
In FCIQMC, as in other QMC approaches, the wave function coefficients $\bs{C}$ are sampled by a collection of walkers. If we define the number of walkers on $| D_j \ket$ as $N_j \in \mathbb{N}$, then the amplitude of each walker can be defined as $C_j / N_j$. A stochastic algorithm to perform the above evolution can then be realized by the following steps:
\begin{enumerate}
\small{
\item \emph{Spawning:} Loop over all occupied determinants, $|D_j\ket$. For each walker on $|D_j\ket$, choose one connected determinant, $|D_i\ket$ ($i \ne j$ and $H_{ij} \ne 0$), with some probability $\Pgen(i \leftarrow j)$. Then create a spawned walker on $|D_i\ket$ with amplitude $ - \Delta \tau \times ( H_{ij} / \Pgen(i \leftarrow j) ) \times ( C_j / N_j ) $.
\item \emph{Death:} Loop over all occupied determinants. Each determinant $|D_i\ket$ spawns to itself with amplitude $ - \Delta \tau ( H_{ii} - E_S ) C_i $.
\item \emph{Annihilation:} Sum together all current and spawned walkers on each occupied determinant to get the new coefficients, $C_i$.
\item \emph{Rounding:} For all determinants with an absolute amplitude, $|C_i|$, less than $1$, stochastically round the absolute amplitude down to $0$ (kill the walker) with probability $1 - |C_i|$, or up to $1$ with probability $|C_i|$.
}
\end{enumerate}
It can be seen that the death step exactly includes the diagonal contribution to $- \Delta \tau \sum_j ( H_{ij} - E_S \delta_{ij} ) C_j$, while the spawning step corresponds to stochastically sampling off-diagonal terms. Rather than looping over all off-diagonal elements in the above summation, precisely one element is chosen for each walker, with some probability $\Pgen(i \leftarrow j)$. The size of the spawned amplitude must then be divided by this probability to keep the algorithm unbiased, so that the average spawned weight is correct.

The shift, $E_S$, is updated slowly to oppose changes in the walker population. This is done every $A$ iterations by
\begin{equation}
E_S(\tau + A\Delta\tau) = E_S(\tau) - \frac{\xi}{A \Delta\tau} \textrm{ln}\left( \frac{N_{\textrm{w}}(\tau+A\Delta\tau)}{N_{\textrm{w}}(\tau)} \right),
\label{eq:shift_update}
\end{equation}
where $\xi$ is a damping parameter, and $N_{\textrm{w}} = \sum_i |C_i|$ is the total walker population.

Note that the above definition of the FCIQMC algorithm uses non-integer walker amplitudes, $C_i$, as first suggested by Umrigar and co-workers\cite{Petruzielo2012}. This differs from the original FCIQMC presentation\cite{Booth2009}, where integer values of $C_i$ were enforced. The use of non-integer coefficients improves the efficiency of the method. In the same work\cite{Petruzielo2012}, Umrigar and co-workers also introduced a semi-stochastic adaptation, in which the projection operator is applied exactly within an important subspace (the \emph{deterministic} or \emph{core} space) and by the above stochastic algorithm otherwise, further reducing stochastic noise.

The energy is commonly estimated by
\begin{align}
\Eref &= \frac{ \bra D_0 | \hat{H} | \Psi \ket }{ \bra D_0 | \Psi \ket }, \\
      &= \frac{ \sum_j H_{0j} C_j }{ C_0 },
\label{eq:hf_estimator}
\end{align}
where the subscript `$0$' refers to the Hartree--Fock determinant or other reference state. A related estimator has been used\cite{Petruzielo2012} where $|D_0\ket$ is replaced by a multi-determinant trial wave function, which again reduces stochastic noise in the estimates.

\subsection{The initiator approximation and walker blooms}
\label{sec:init_approx}

The above algorithm allows the exact FCI wave function to be sampled without bias. However, in practice a population plateau appears in the simulation, below which the fermion sign problem leads to uncontrollable noise\cite{Spencer2012}. This plateau height therefore sets a minimum memory requirement on the simulation, which is typically much smaller than that required to store the FCI space, but which nonetheless grows exponentially with the system size. As such, the FCIQMC algorithm as stated above is still restricted to small systems.

To overcome this, Cleland \emph{et al.} introduced the initiator approximation to FCIQMC, known as i-FCIQMC\cite{Cleland2010, Cleland2011}. In this, all determinants with a weight greater than $n_a$ are defined as initiators (with $n_a$ equal to $2$ or $3$, typically). Initiators are allowed to spawn to any determinant, while non-initiators may only spawn to already-occupied determinants. Attempted spawnings from non-initiators to unoccupied determinants are removed from the simulation. An exception occurs if two non-initiators spawn to the same determinant in the same iteration, in which case the spawnings are allowed (the `coherent spawning rule'). When the semi-stochastic adaptation\cite{Petruzielo2012, Blunt2015} is used, all determinants within the deterministic space are also made initiators. We note that in some cases this deterministic space may be large, in which case the initiator error can change significantly.

This initiator approach significantly reduces the sign problem in the method, allowing arbitrarily-small walker populations to be used. In exchange, an approximation is introduced, as the Hamiltonian is effectively truncated; the above approximation is equivalent to setting Hamiltonian elements between non-initiators and unoccupied determinants to zero. As the walker population is increased, the number of initiators and occupied determinants both increase, and i-FCIQMC tends towards the exact solution. Therefore, i-FCIQMC provides a systematic way to converge to the FCI limit.

An important concept in i-FCIQMC is that of a ``walker bloom''. A bloom is defined as a spawning event with weight greater than $n_a$, such that the new determinant instantly becomes an initiator. Such events should be avoided, as they lead to essentially random determinants being made initiators. Furthermore, with enough bloom events we find that the initiator space grows exponentially in $\tau$, and a sign problem returns. This criterion is often used to set the time step, $\Delta \tau$, which is chosen so as to prevent bloom events (or to allow only a small number to occur each iteration).

\section{FCIQMC with preconditioning}
\label{sec:precond_fciqmc}

\subsection{Algorithm definition}

Imaginary-time evolution as described in Section~\ref{sec:fciqmc} will converge to the ground state of $\hat{H}$ only if $\Delta \tau$ is chosen to obey
\begin{equation}
\Delta \tau < \frac{2}{E_{\textrm{max}} - E_0},
\end{equation}
where $E_{\textrm{max}}$ and $E_0$ are the highest and lowest energy eigenvalues of $\hat{H}$, respectively. For large systems and basis sets, we find that this condition restricts the time step to be of order $\Delta \tau \sim 10^{-3}$ au, or even smaller. Typically, FCIQMC may take on the order of $\sim 10^{3} - 10^{5}$ iterations for the initial transient to decay, allowing sampling of the ground state to begin.

Preconditioning is a commonly-used approach to speed up the iterative solution of a system of linear equations\cite{axelsson1994, Beauwens2004, Saad2011}. For an eigenvalue problem $\bs{H} \bs{C} = E \bs{C}$, an iterative solution may be obtained by
\begin{equation}
\bs{C}_{n+1} = \bs{C}_n - \gamma_n \bs{P}^{-1} ( \bs{H} \bs{C}_n - E_n \bs{C}_n ),
\end{equation}
where $\bs{P}$ (or often $\bs{P}^{-1}$) is referred to as the preconditioner, $\bs{C}_n$ and $E_n$ are best estimates of $\bs{C}$ and $E$ from iteration $n$, and $\gamma_n$ is a step size. Setting $\bs{P} = \bs{I}$ and $\gamma_n = \Delta \tau$ returns imaginary-time propagation as in FCIQMC. However, for an appropriate choice of $\bs{P}^{-1}$, convergence can be sped up considerably. The most common choice is the Jacobi preconditioner, defined as $P_{ij} = (H_{ii} -E)\delta_{ij}$, which is widely used throughout quantum chemistry, such as in the Davidson method\cite{Saad2011}. It should also be noted that the update coefficients are essentially equal to those obtained through first-order perturbation theory.

We therefore suggest the following update equation for the FCIQMC:
\begin{equation}
C_i(\tau + \Delta \tau) = C_i(\tau) - \frac{\Delta \tau}{H_{ii} - E} \sum_{j} ( H_{ij} - E \delta_{ij} ) C_j(\tau).
\label{eq:precond_2}
\end{equation}
For consistency, we have again used $\Delta \tau$ to denote the step size. However, it should be emphasized that taking the limit $\Delta \tau \to 0$ does not result in the imaginary-time Schr{\"o}dinger equation, and equal values of $\Delta \tau$ do not give equal rates of convergence with and without preconditioning, so care should be taken in comparisons.

Exactly as has been done for FCIQMC with imaginary-time propagation, it is simple to write down an FCIQMC algorithm for the above preconditioned evolution:
\begin{enumerate}
\item \emph{Spawning:} Loop over all occupied determinants, $|D_j\ket$, and for each walker perform $N_{\textrm{spawn}}$ spawning attempts. For each spawning attempt from $|D_j\ket$, choose one connected determinant, $|D_i\ket$ ($i \ne j$ and $H_{ij} \ne 0$), with some probability $\Pgen(i \leftarrow j)$. Then create a spawned walker on $|D_i\ket$ with amplitude $ - \Delta \tau \times ( H_{ij} / \Pgen(i \leftarrow j) ) \times ( C_j / N_{\textrm{spawn}} N_j ) $.
\item \emph{Apply the preconditioner to spawnings:} Loop over determinants to which spawnings have occured. For a spawned walker on $|D_i\ket$, multiply its amplitude by $1/(H_{ii} - E)$.
\item \emph{Death:} Loop over all occupied determinants. Multiply each determinant's amplitude by $1 - \Delta \tau$.
\item \emph{Annihilation:} Sum together all current and spawned walkers on each occupied determinant to get the new coefficients, $C_i$.
\item \emph{Rounding:} For all determinants with an absolute amplitude, $|C_i|$, less than $1$, stochastically round the absolute amplitude down to $0$ (kill the walker) with probability $1 - |C_i|$, or up to $1$ with probability $|C_i|$.
\end{enumerate}
Most of the algorithm is the same as for FCIQMC with imaginary-time evolution, and we will compare the two algorithms in Section~\ref{sec:comparison}. In particular, the annihilation and rounding steps are identical. The main differences are that spawned walkers now have the preconditioner $1/(H_{ii} - E)$ applied, and the death step is also appropriately modified (simplified, in fact) to account for this same factor. We have also chosen to allow each walker to make $\Nspawn$ spawning attempts, so that the amplitude of each spawned walker must be divided by the same factor to keep the algorithm unbiased. Here, $\Nspawn$ is some integer equal to $1$ or greater. In previous applications of FCIQMC, this has always been taken as $\Nspawn = 1$.

One may ask precisely when the evolution of Eq.~(\ref{eq:precond_2}) converges. Fixed points of this evolution are when $\bs{H} \bs{C} - E \bs{C} = \bs{0}$, as desired. Theoretically, convergence is guaranteed provided the iteration matrix has a spectral radius less than $1$, which is met for diagonally-dominant matrices (although this is not necessary). In practice, the proposed evolution is well established as being extremely successful. We have tested this for a range of systems, including both molecular and model systems with both weak and strong correlation, and have always found convergence to occur for $\Delta \tau = 0.5$ au or smaller, and often with $\Delta \tau = 1.0$ au.

The use of much larger time steps has an important consequence, which must be emphasized: the size of each spawned walker is proportional to $\Delta \tau$, so that larger spawned walkers will be created with larger time steps. Having very large spawning events (i.e., larger than $4$ or so) can significantly increase the stochastic noise in a simulation. The use of multiple spawning attempts per walker ($\Nspawn > 1$) was introduced above as a way to counter this. The size of each spawned walker will be proportional to $\Delta \tau / \Nspawn$, giving a way to reduce the maximum spawning size by increasing $\Nspawn$. This will increase the cost per iteration, therefore reducing the savings of using a large $\Delta \tau$, a point which we will return to.

We typically choose to initialize the wave function using configuration interaction singles and doubles (CISD), which is always feasible for systems currently amenable to FCIQMC. However, this is not required in general.

We note that this preconditioned approach is similar to a previous modification to FCIQMC and coupled cluster Monte Carlo\cite{Thom2012, Franklin2016}, changing the step taken by a quasi-Newton approach\cite{NeufeldThom_unpublished}, which though implemented\cite{HANDEv1.1,HANDE} has yet to be widely used. An alternative approach within a deterministic framework was considered by Zhang and Evangelista\cite{Zhang2016}, who considered a Chebyshev expansion of the exponential propagator.

\subsection{Population control: intermediate normalization}

As described in Section~\ref{sec:fciqmc}, with imaginary-time evolution the walker population is typically controlled by a shift, $E_S(\tau)$, which is updated by Eq.~(\ref{eq:shift_update}).

With preconditioning, a more natural choice is intermediate normalization. Consider the projected energy estimator, $\Eref$, defined in Eq.~(\ref{eq:hf_estimator}), which is equally valid both with and without preconditioning. If we set the energy in the preconditoner to equal this estimate ($E = \Eref$) then it can be seen that $\sum_{j} ( H_{0j} - E \delta_{0j} ) C_j = 0$. If evolving with Eq.~(\ref{eq:precond_2}), it is then simple to check that the coefficient $C_0$ on $| D_0 \ket$ remains exactly constant throughout. We note that $|D_0\ket$ need not be the Hartree--Fock determinant, and can also be updated during a simulation to match the most populated determinant. This choice of population control does not restrict the method to weakly correlated systems.

For $C_0$ to remain exactly constant, the estimate $\sum_{j} H_{0j} C_j$ must be obtained from the spawnings made to $|D_0\ket$ from the latest iteration. It is helpful to use a deterministic space containing $|D_0 \ket$ and its most important connections, to avoid the situation where no spawnings are made to $|D_0 \ket$ in an iteration.

With this choice of population control, the walker population will grow in the early iterations of the simulation, settling down and fluctuating about a final value once convergence has been achieved. This makes choosing a final walker population more difficult than in the original scheme. However, this can usually be achieved by performing a preliminary test with a small initial population, and then scaling appropriately.

The above modification has an effect on the projected energy estimator, defined in Eq.~(\ref{eq:hf_estimator}), which should be noted: the population $C_0$ now remains exactly constant, and so is not a random variable. Typically in FCIQMC, one would average the numerator and denominator of Eq.~(\ref{eq:hf_estimator}) separately, and perform the required division \emph{after} this averaging. Now that the denominator is constant, this separate averaging makes no difference. This deserves consideration, as in the original approach this estimator can theoretically be biased if performed as $\langle x/y \rangle$ rather than $\langle x \rangle / \langle y \rangle$ (although any such bias is essentially negligible, in our experience). Does this preconditioned approach remove all such bias? We suspect that the answer is ``no'', and that this theoretical bias is transferred to the sampling of $| \Psi(\tau) \ket$, due to the applications of $1/(H_{ii} - E)$ in the propagation (where $E$ is a random variable), and population control bias\cite{Vigor2015} due to the aggressive updates to $E$. However, we emphasize that any such bias seems to be essentially negligible in practice.

We note that this intermediate normalization approach has recently been used in a related QMC approach to coupled cluster theory\cite{Scott2019}. A related approach to population control has also been used recently by Alavi and co-workers in FCIQMC with imaginary-time propagation\cite{fixed_n0_unpublished}.

\begin{figure*}[t]
\includegraphics{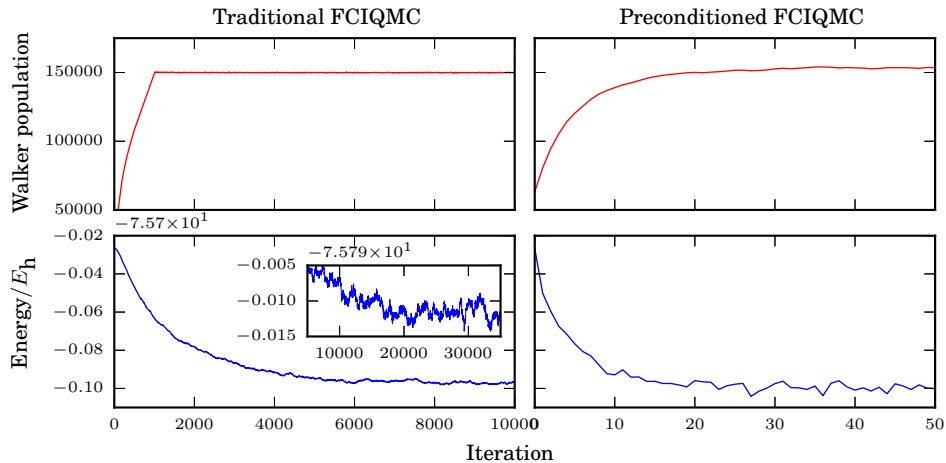}
\caption{An example comparison of convergence between the traditional and preconditioned FCIQMC approaches. The system is C$_2$ in a cc-pVQZ basis set, at equilibrium geometry. Without preconditioning, $\Nspawn=1$ and $\Delta \tau = 8 \times 10^{-4}$ au. With preconditioning, $\Nspawn=200$ and $\Delta \tau = 0.4$ au. It must be emphasized that each iteration in the preconditioned approach is $\sim 200$ times more expensive than in the traditional method, because of the difference in $\Nspawn$.}
\label{fig:converge_comp}
\end{figure*}

\subsection{The initiator approximation}
\label{sec:precond_init}

In preconditioned FCIQMC, the initiator adaptation is largely unchanged: initiators are defined as determinants with an absolute amplitude $|C_i|$ greater than $n_a$, which is set to $2$ or $3$, typically. Attempted spawnings from initiators are always accepted, but spawnings from non-initiators are only accepted if made to already-occupied determinants, else they are removed from the simulation. We again use the semi-stochastic adaptation, where all deterministic states are also defined as initiators.

We again emphasize the importance of avoiding bloom events in the initiator approximation. Given that $\Delta \tau$ can be made much larger compared to the original FCIQMC approach, it is important then to increase $\Nspawn$ appropriately in order to avoid walker blooms. Avoiding these large spawning events is important in any QMC approach to control statistical noise, but is perhaps particularly important in the initiator adaptation, where such blooms lead to random determinants being given initiator status.

Lastly, we note that with large $\Nspawn$ it is necessary to remove the `coherent spawning' rule of the i-FCIQMC. That is, we do \emph{not} allow simultaneous spawnings to an unoccupied determinant from two non-initiators to survive. For large $\Nspawn$, such events become a frequent occurrence, and we often encounter a sign problem re-emerging. Removing this rule has only a very small effect on the accuracy of the initiator approximation.

\section{Comparison of FCIQMC with and without preconditioning}
\label{sec:comparison}

\subsection{Algorithm}

Here we state the differences between the original and preconditioned FCIQMC approaches. Specifically, changes relative to the original FCIQMC algorithm are:
\begin{enumerate}
\item Once spawned walkers have been generated, the preconditioner $1/(H_{ii} - E)$ must be applied to each.
\item In the death step, a factor of $1 - \Delta \tau$ is applied to each walker coefficient, rather than shifting each coefficient by $ - \Delta \tau ( H_{ii} - E_S ) C_i $
\item The shift $E_S$ is replaced by a separate energy estimate, which is obtained from Eq.~(\ref{eq:hf_estimator}). This energy is not needed in the death step, but is instead required in the preconditioner.
\item In order to reduce the size of spawned walkers in the presence of a very large $\Delta \tau$, we allow each walker to make $\Nspawn$ spawning attempts. In the original approach, each walker only makes $1$ spawning attempt (i.e., $\Nspawn = 1$).
\end{enumerate}

\subsection{Implementation}

The use of preconditioning does not significantly change the implementation of FCIQMC, and only a few changes may be required to implement preconditioning in an existing FCIQMC code. In particular, all communication of spawned walkers is performed in the same manner. In both approaches, spawned walkers are held in a separate array to the current walkers. These spawned walkers are communicated to their parent process and annihilated to give a final merged spawning array, which we denote $\bs{S}$. This spawning array may then be annihilated with the main walker list, $\bs{C}$, to give the new walker coefficients. One can write down an expression for the expectation value of the spawning array $\bs{S}$,
\begin{align}
S_i &= - \Delta \tau \bra D_i | \hat{H}_{\textrm{off}} | \Psi \ket \\
S_i &= - \Delta \tau \sum_{j \ne i} H_{ij} C_j,
\label{eq:spawn_array}
\end{align}
where $\hat{H}_{\textrm{off}}$ contains only off-diagonal elements in the basis set used. Importantly, $\bs{S}$ only contains off-diagonal elements of $\hat{H}$ because diagonal elements are accounted for separately by the death step. Note also that we have dropped the $\tau$ dependence in $\bs{S}(\tau)$ and $|\Psi(\tau)\ket$ for notational clarity later. This identification of the spawning array will be crucial in Section~\ref{sec:estimators_theory} for constructing more efficient energy estimators in FCIQMC.

In the preconditioned case, the factors of $1/(H_{ii} - E)$ are then applied directly to the spawning array $\bs{S}$ after it has been communicated and merged across processors, but before it is merged with the previous walker, list $\bs{C}$.

The diagonal Hamiltonian element $H_{ii}$ can be calculated for the new determinant $|D_i\ket$ in $\mathcal{O}(N)$ time from the value $H_{jj}$ of the parent walker on $|D_j\ket$, which is always stored. Therefore, it is \emph{not} required to perform a full $\mathcal{O}(N^2)$ construction of $H_{ii}$ for each spawned walker, which would be expensive.

\subsection{An example: C$_2$ cc-pVQZ}

As a simple demonstration, in Fig.~(\ref{fig:converge_comp}) we compare the convergence of C$_2$ at equilibrium bond length, in a cc-pVQZ basis and with a frozen core, both with and without preconditioning. In both cases the simulation is initialized from the CISD wave function. For FCIQMC without preconditioning, we choose $\Nspawn=1$, and set the time step so as to prevent bloom events (giving $\Delta \tau = 8 \times 10^{-4}$ au), which is the standard protocol in most current FCIQMC calculations. For FCIQMC with preconditioning, we first choose a time step of $\Delta \tau = 0.4$ au and then choose $\Nspawn=200$ so as to prevent bloom events. Note that the energy estimator used here is the projected energy estimator, $\Eref$, defined in Eq.~(\ref{eq:hf_estimator}).

It can be seen that, while FCIQMC without preconditioning requires $\sim 3 \times 10^4$ iterations to fully converge, convergence with preconditioning is achieved within $30$ iterations. It is very important to emphasize, however, that the iteration time is roughly proportional to $\Nspawn$. Therefore, each iteration with $\Nspawn=200$ is roughly $200$ times more expensive than without. Even with this taken into account, convergence is quicker with preconditioning than without, at least in this case. More careful comparison and discussion is given in Section~\ref{sec:convergence}.

\section{Improved estimators in FCIQMC}
\label{sec:estimators_theory}

Separately from the above discussion of preconditioning in FCIQMC, we now discuss the calculation of improved energy estimators in FCIQMC, including perturbative corrections to initiator error. We emphasize that all of the following estimators can be calculated identically in both the original and preconditioned FCIQMC approaches. Although the following section is separate from the previous section on preconditioning in FCIQMC, we will show in Section~\ref{sec:error_reduction} that the preconditioned approach greatly benefits the calculation of PT2-based estimators in FCIQMC, and so we will ultimately be interested in their application together.

\subsection{Sampling variational energies without reduced density matrices}

In addition to the projected energy estimators (comprising both projections onto single determinants and multi-determinant trial solutions), variational energy estimators have also been used in FCIQMC\cite{Overy2014, Blunt2017}:
\begin{equation}
\Evar = \frac{ \bra \Psi | \hat{H} | \Psi \ket }{ \bra \Psi | \Psi \ket }.
\end{equation}
Consider the numerator. Because $| \Psi \ket$ is a stochastic estimate, the replica trick must be used to ensure that this estimator is unbiased, as has been described elsewhere\cite{Zhang1993, Overy2014, Blunt2014, Blunt2015_2}. In this, two independent FCIQMC simulations are performed, which we label $ | \Psi^1 \ket = \sum_i C_i^1 |D_i \ket $ and $| \Psi^2 \ket = \sum_i C_i^2 |D_i \ket $. Then the variational estimate can be obtained as
\begin{equation}
\Evar = \frac{ \big\langle \; \sum_{ij} C_i^1 H_{ij} C_j^2 \; \big\rangle }{ \big\langle \; \sum_i C_i^1 C_i^2 \; \big\rangle },
\label{eq:var_energy_def}
\end{equation}
where $\big\langle \ldots \big\rangle$ denotes an average over the simulation after convergence, and we assume real coefficients throughout. We drop this averaging notation for clarity, but it should be understood that the numerator and denominator of each estimator is averaged (separately) over the simulation from convergence onwards.

In previous FCIQMC studies, Eq.~(\ref{eq:var_energy_def}) has been calculated as $\textrm{Tr}(\hat{\Gamma} \hat{H})$, where $\hat{\Gamma}$ is the two-particle density matrix (2-RDM), whose calculation in FCIQMC was described in Refs.~(\onlinecite{Overy2014}) and (\onlinecite{Blunt2017}). The efficient implementation of 2-RDMs in FCIQMC is involved, and their accumulation can slow the simulation down by a significant factor. In this study we therefore calculate $\Evar$ and related quantities directly.

The two main large arrays in an FCIQMC implementation are $\bs{C}$ (with components $C_i^r = \bra D_i | \Psi^r \ket$) and $\bs{S}$ (with components $S_i^r = -\Delta \tau \sum_{j \ne i} H_{ij} C_j^r$) as defined already. $\bs{S}$ is distributed across processes with the same mapping as $\bs{C}$, such that it is easy to take a dot product between $\bs{C}$ and $\bs{S}$. In the following, we therefore write all estimators in terms of $\bs{C}$ and $\bs{S}$, showing how they are efficiently calculated in practice. Again we emphasize that these arrays are constructed in the same manner for both original and preconditioned algorithms, so that all of the following applies for both approaches. In the preconditioned case, the preconditioner is applied to $\bs{S}$ only \emph{after} the following estimators are constructed.

$\Evar$ may be calculated as
\begin{align}
\Evar &= \frac{ \sum_i C_i^1 [ \sum_{j \ne i} H_{ij} C_j^2 + H_{ii} C_i^2 ] }{\sum_i C_i^1 C_i^2 }, \\
                 &= \frac{ \sum_i C_i^1 [ -S_i^2 / \Delta \tau + H_{ii} C_i^2 ] }{ \sum_i C_i^1 C_i^2 },
\end{align}
and statistical errors can be reduced by making use of spawnings from both replicas:
\begin{equation}
\Evar = \frac{ \sum_i C_i^1 H_{ii} C_i^2 }{ \sum_i C_i^1 C_i^2 } - \frac{1}{2 \Delta \tau} \frac{ \sum_i [ C_i^1 S_i^2 + S_i^1 C_i^2 ] }{ \sum_i C_i^1 C_i^2 }.
\label{eq:var_energy}
\end{equation}
Note that only the expectation value of this estimator is variational. Instantaneous estimates are not.

\subsection{Perturbative corrections to initiator error}
\label{sec:pt2_estimators}

Given that the variational energy estimator $\Evar$ is based on an inexact wave function subject to the initiator approximation, we recently suggested\cite{Blunt2018} a second-order perturbative correction to this estimator
\begin{equation}
\Delta E_2 = \frac{1}{(\Delta\tau)^2} \sum_a \frac{ S^1_a S^2_a }{ E - H_{aa} },
\label{eq:en2_original}
\end{equation}
where the summation is performed over all spawnings which are cancelled due to the initiator criterion, and there is a normalization factor of $\bra \Psi^1 | \Psi^2 \ket$. From this, a total energy estimate can be defined as
\begin{equation}
\EvarPT = \Evar + \Delta E_2.
\label{eq:evar_pt}
\end{equation}
The above formula for $\Delta E_2$ was constructed by analogy with SCI+PT2, where configuration interaction is performed within a truncated space, beyond which a second-order Epstein-Nesbet perturbative correction can be constructed. Initiator FCIQMC can also be loosely seen as a truncated method, which allows the above estimator to be written down by analogy, making use of spawned walkers which are otherwise thrown away without use. However, a truncated space for i-FCIQMC is somewhat poorly defined, first because the space of occupied and initiator determinants is non-constant, and second because some unoccupied determinants are connected to both initiators and non-initiators (as such, the truncation is more precisely on the Hamiltonian, not the space).

To make this perturbative correction more rigorous, we consider a slightly different estimator, which we call $\Ep$, which can then be compared to $\EvarPT$ for any deviations. Given the wave function within the initiator approximation, $|\Psi\ket$, it is possible to write down a more accurate wave function which we denote $|\Phi\ket = \sum_i \Phi_i |D_i\ket$,
\begin{equation}
| \Phi \ket = [ E - \hat{H}_{\textrm{d}} ]^{-1} \hat{H}_{\textrm{off}} | \Psi \ket,
\label{eq:projected_wf}
\end{equation}
\begin{equation}
\Phi_i = \frac{1}{ E - H_{ii} } \sum_{j \ne i} H_{ij} C_j,
\label{eq:projected_wf_comp}
\end{equation}
where $\hat{H}_{\textrm{d}} = \sum_i H_{ii} |D_i \ket \bra D_i|$. An energy estimator based upon this improved wave function can then be written down as
\begin{align}
\Ep &= \frac{ \bra \Phi | \hat{H} | \Psi \ket }{ \bra \Phi | \Psi \ket  }, \\
                      &= \frac{ \bra \Psi | \hat{H}_{\textrm{off}} \; [ E - \hat{H}_{\textrm{d}} ]^{-1} \hat{H} | \Psi \ket }{ \bra \Psi | \hat{H}_{\textrm{off}} \; [ E - \hat{H}_{\textrm{d}} ]^{-1} | \Psi \ket  }.
\end{align}
This expression can be expanded in terms of $\bs{C}$ and $\bs{S}$ components to give an estimator for use in FCIQMC. First the numerator,
\begin{align}
\bra \Phi^1 | \hat{H} | \Psi^2 \ket &= \sum_i \frac{1}{E - H_{ii}} \sum_{k \ne i} C_k^1 H_{ki} \sum_j H_{ij} C_j^2, \\
                                    &= \sum_i \frac{ [ -S_i^1 / \Delta\tau] [ -S_i^2/\Delta\tau + H_{ii}C_i^2] }{E - H_{ii}}, \\
                                    &= \frac{1}{(\Delta\tau)^2} \sum_i \frac{ S_i^1 S_i^2 }{ E - H_{ii} } - \frac{1}{\Delta\tau} \sum_i \frac{S_i^1 H_{ii} C_i^2 }{ E - H_{ii} },
\label{eq:precond_energy}
\end{align}
and similarly the denominator by
\begin{align}
\bra \Phi^1 | \Psi^2 \ket &= \sum_i \frac{ \sum_{j \ne i} C_j^1 H_{ji} C_i^2 }{ E - H_{ii} }, \\
                          &= \frac{-1}{\Delta\tau} \sum_i \frac{ S_i^1 C_i^2 }{ E - H_{ii} }.
\label{eq:precond_denom}
\end{align}
As for $\Evar$, the cross terms including $C_i$ and $S_i$ can be averaged with both combinations of replicas $1$ and $2$, both in the numerator and denominator.

It can be seen that all of the terms in $\EvarPT$ are also included in $\Ep$. The connection of $\Ep$ with perturbation theory is made precise in Appendix~(\ref{sec:appendix}). However, a simple way to see this connection is to note that $| \Phi \ket$ can be expressed as $| \Psi_0 \ket + | \Psi_1 \ket$, where $ | \Psi_1 \ket $ is the first-order Epstein-Nesbet correction to an appropriate zeroth-order wave function, $ | \Psi_0 \ket $. It is then simple to show that $\Ep$ includes a second-order perturbative correction.

The estimator $\Ep$ has the advantage that is requires no partitioning between a variational and non-variational space.  Furthermore, $\Ep$ takes the form $\bra \Phi | \hat{H} | \Psi \ket / \bra \Phi | \Psi \ket$, where $|\Psi\ket$ and $|\Phi\ket$ are both wave functions accessible from FCIQMC. This is the form of a traditional estimator in a QMC method, and avoids explicitly adding a perturbative correction. On the other hand $\EvarPT$ usually has smaller statistical noise, and so is often more useful in practice.

Note that in Eq.~(\ref{eq:en2_original}), (\ref{eq:precond_energy}) and (\ref{eq:precond_denom}), we calculate $E$ using the projected energy estimator, $\Eref$. We could also use $\Evar$, but find that this makes little difference in practice [as $E$ is well separated from any $H_{ii}$, particularly for Eq.~(\ref{eq:en2_original})].

\subsection{Sampling the variance of the energy}

Finally, we point out that it is simple to write the energy variance, $\sigma^2$, as efficient operations involving $\bs{C}$ and $\bs{S}$, and therefore to sample in FCIQMC. Ignoring normalization,
\begin{equation}
\sigma^2 = \bra \Psi | \hat{H}^2 | \Psi \ket - \bra \Psi | \hat{H} | \Psi \ket^2.
\label{eq:variance}
\end{equation}
We emphasize that this is the standard energy variance, and not some measure of statistical error. The calculation of $\bra \Psi | \hat{H} | \Psi \ket$ has been discussed already. The calculation of $\bra \Psi | \hat{H}^2 | \Psi \ket$ is performed (using replica sampling) as:
\begin{align}
\bra \Psi^1 | \hat{H}^2 | \Psi^2 \ket &= \sum_{ijk} C_i^1 H_{ij} H_{jk} C_k^2, \\
                                      &= \sum_j \sum_i C_i^1 H_{ij} \sum_k H_{jk} C_k^2, \\
                                      &= \sum_j \Big( [ C_j^1 H_{jj} + \sum_{i \ne j} C_i^1 H_{ij} ] \nonumber \\
                                      &\;\;\;\;\;\;\;\; \times [ H_{jj} C_j^2 + \sum_{k \ne j} H_{jk} C_k^2 ] \Big), \\
                                      &= \sum_j C_j^1 H_{jj}^2 C_j^2 \nonumber \\
                                       &\;\;\; - \frac{1}{\Delta\tau} \sum_j [ C_j^1 H_{jj} S_j^2 + S_j^1 H_{jj} C_j^2 ] \nonumber \\
                                       &\;\;\; + \frac{1}{\Delta\tau^2} \sum_j S_j^1 S_j^2.
\label{eq:h_squared}
\end{align}
Note that the expression for $\sigma^2$ involves squaring the estimate of $\bra \Psi | \hat{H} | \Psi \ket$. However, this operation can be performed \emph{after} averaging over the simulation, such that bias is not a concern here.

The energy variance could be useful as a measure of initiator error in i-FCIQMC. It could also be used to calculate improved excitation energies in i-FCIQMC by variance matching\cite{Robinson2017}. In previous applications of excited-state FCIQMC\cite{Blunt2015_3}, the same walker	population was used for ground and excited states. Since excited states require larger walker populations for similar accuracy, this leads to an imbalance in accuracy between the two states. We expect that variance matching could improve this situation, and could perhaps also benefit model space QMC\cite{Ten-no2013, Ohtsuka2015} in the same way.

\begin{figure}[t]
\includegraphics{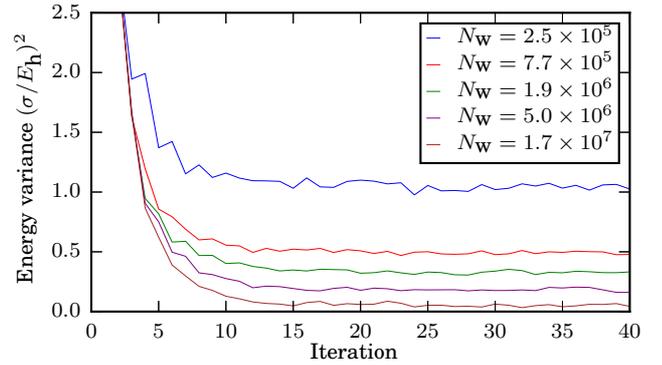}
\caption{The variance of the energy ($\sigma^2$) from FCIQMC for the Hubbard model with $U/t=4$, for a periodic, two-dimensional $18$-site lattice at half-filling. Initiator error is converged by increasing $\Nw$. Preconditioning was used with a time step of $\Delta \tau = 1.0$ au. The stated populations ($\Nw$) are the final average values per replica. As expected, $\sigma^2$ tends to $0$ as initiator error is removed.}
\label{fig:variance}
\end{figure}

Figure~(\ref{fig:variance}) shows convergence of $\sigma^2$ with iteration number (with preconditioning and $\Delta \tau = 1.0$ au) and walker population per replica ($\Nw$) for the Hubbard model at $U/t=4$, on a periodic two-dimensional $18$-site lattice at half-filling. The lattice is the same as that presented in Supplemental Material of Ref.~(\onlinecite{Blunt2015_2}). As expected, $\sigma^2$ tends to $0$ as the walker population is increased and initiator error removed.

\section{Results}
\label{sec:results}

\begin{table*}[t]
\begin{center}
{\footnotesize
\begin{tabular}{@{\extracolsep{4pt}}lcccccc@{}}
\hline
\hline

 & & \multicolumn{1}{c}{$\Evar$} & \multicolumn{2}{c}{$\EvarPT$} & \multicolumn{2}{c}{$\Ep$} \\
\cline{3-3} \cline{4-5} \cline{6-7}
System & $N_{\textrm{w}}$ & Error/$\textrm{m}\Eh$ & Error/$\textrm{m}{\Eh}$ & \% corrected & Error/$\textrm{m}{\Eh}$ & \% corrected \\
\hline
C$_2$ (equilibrium, cc-pVQZ)               & $1.75 \times 10^5$ & 2.20(5)  & 0.10(5)  & 95(4)   & 0.05(5)  & 97(4)   \\
C$_2$ (stretched, cc-pVQZ)                 & $1.23 \times 10^5$ & 3.0(1)   & 0.3(1)   & 89(6)   & 0.5(1)   & 82(5)   \\
Formaldehyde (aug-cc-pVDZ)                 & $3.0  \times 10^5$ & 4.0(1)   & 0.3(1)   & 93(4)   & 0.02(22) & 100(6)  \\
Formamide (cc-pVDZ)                        & $4.8  \times 10^6$ & 7.2(3)   & 0.8(3)   & 112(8)  & 0.6(4)   & 108(8)  \\
Butadiene $(22\textrm{e},82\textrm{o})^a$  & $8.8  \times 10^7$ & 12.9(4)  & 0.4(7)   & 97(6)   & 1.0(10)  & 92(8)   \\
Hubbard model ($U/t=2$)                    & $1.1  \times 10^4$ & 4.6(1)   & 0.6(1)   & 87(4)   & 0.51(5)  & 89(3)   \\
Hubbard model ($U/t=4$)                    & $2.5  \times 10^5$ & 66.49(9) & 23.73(9) & 64.3(2) & 23.7(1)  & 64.3(2) \\
\hline
\hline
\end{tabular}
}
\caption{Example improvements of $\EvarPT$ and $\Ep$ relative to $\Evar$, for a variety of systems. The frozen-core approximation and Hartree--Fock orbitals are used for molecular systems. The walker population is chosen deliberately small so that there is substantial initiator error in $\Evar$. $\EvarPT$ and $\Ep$ have almost identical accuracy, but $\EvarPT$ typically has smaller noise. Hubbard model calculations are performed at half filling on an $18$-site lattice, and errors here are calculated relative to FCI values. Errors for other systems are calculated relative to very accurate extrapolated benchmarks (see the main text for details). Equilibrium and stretched nuclear distances for C$_2$ are $R=1.24253$ \AA and $R=2.0$ \AA, respectively. ${}^a$The basis set for butadiene is ANO-L-VDZP$[3s2p1d]/[2s1p]$, as used previously\cite{Daday2012,Olivares2015,Chien2018,Guo2018_1}.}
\label{tab:examples}
\end{center}
\end{table*}

The results are structured as follows. Example results for the perturbatively-corrected estimators are presented in Section~\ref{sec:examples}. In Section~\ref{sec:error_reduction} it is shown that the efficiency of such estimators is greatly increased by performing multiple spawning attempts per walker (large $\Nspawn$). The effect of correlation of QMC data on perturbative corrections is discussed in Section~\ref{sec:corr_length}, and the convergence time of FCIQMC with preconditioning is considered in Section~\ref{sec:convergence}. Finally, we show application to a larger example, benzene.

All molecular geometries are presented in supporting information. The geometry of formamide and benzene were taken from Ref.~(\onlinecite{Schreiber2008}). The geometry for butadiene was taken from Ref.~(\onlinecite{Daday2012}).

The initiator threshold $n_a$ was set to $3.0$ for all systems except for C$_2$, where it was set to $2.0$ (for consistency with results in Ref.~[\onlinecite{Blunt2015_3}]).

The preconditioned approach was implemented in NECI\cite{NECI_github}, which was used for all FCIQMC results. Integral files were generated with PySCF\cite{pyscf}. CC benchmarks were obtained with MRCC\cite{mrcc, Bomble2005, Kallay2005}. SCI+PT2 benchmarks were obtained using the SHCI approach\cite{Holmes2016_2, Sharma2017} with Dice\cite{Dice}.

\subsection{Results for perturbative corrections to initiator error}
\label{sec:examples}

Table~\ref{tab:examples} shows examples of the correction made by $\EvarPT$ and $\Ep$ relative to $\Evar$, for a variety of systems. Walker populations are chosen so that substantial initiator error exists in $\Evar$. Hubbard model calculations are performed at half-filling on the same lattice as used in Fig.~(\ref{fig:variance}). Hartree--Fock orbitals were used for molecular systems. Each error is calculated relative to either the exact FCI energy (for Hubbard model examples) or a very accurate extrapolated estimate (for molecular examples). Benchmarks for C$_2$ are extrapolated SCI+PT2 values from Ref.~(\onlinecite{Holmes2017}). We also obtained benchmarks for formaldehyde and formamide using extrapolated SCI+PT2 (for formamide, these SCI+PT2 calculations used orbitals optimized by performing active-active rotations in an SHCI calculation with a threshold of $\epsilon = 2 \times 10^{-4}$, as described in Ref.~[\onlinecite{Smith2017}]). The benchmark for butadiene is an extrapolated DMRG+PT2 result of $-155.557567$ $\Eh$ from Ref.~(\onlinecite{Guo2018_1}).

The molecular systems considered are weakly correlated and so the PT2 correction is expected to be effective, which is found to be the case. The correction here is typically $> 85\%$, as was found in Ref.~(\onlinecite{Blunt2018}). The correction is less effective for the Hubbard model as the coupling strength is increased.

\begin{figure}[t]
\includegraphics{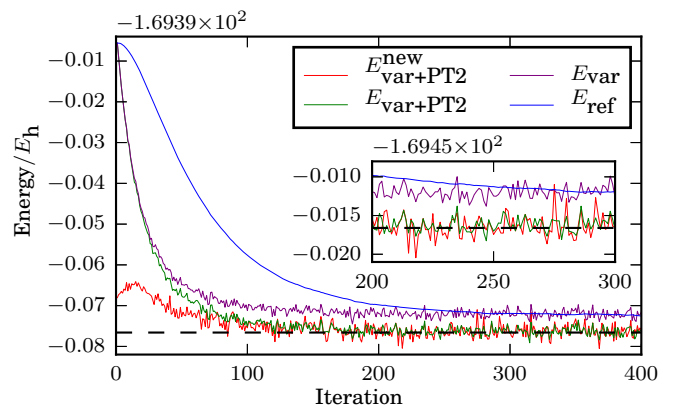}
\caption{Convergence of FCIQMC for formamide, in a cc-pVDZ basis with a frozen core $(18\textrm{e}, 54\textrm{o})$. The FCIQMC wave function is initialized from the CISD wave function, and parameters of $\Delta \tau = 0.03$ au and $60$ spawns per walker were used. The dashed line is an extrapolated SCI benchmark (obtained using optimized orbitals\cite{Smith2017}), which is essentially exact. $\Eref$, $\Evar$ and $\EvarPT$ all begin from the CISD energy, while $\Ep$ is substantially more accurate before any further convergence. $\Eref$, as usually used in FCIQMC, converges slower than other estimators. $\EvarPT$ and $\Ep$ converge to the same value, although $\Ep$ has larger noise. These trends are seen across all systems.}
\label{fig:formamide_converge}
\end{figure}

Results for $\EvarPT$ and $\Ep$ are seen to be essentially identical within error bars. This is expected for the reasons discussed in Section~\ref{sec:pt2_estimators}. However, the statistical error on $\EvarPT$ is usually smaller than that on $\Ep$, so that $\EvarPT$ is generally preferable (although some exceptions occur, particularly for model systems, as seen for the Hubbard model at $U/t=2$ in Table~\ref{tab:examples}). $\EvarPT$ has the disadvantage that its derivation involves a somewhat poorly-defined definition of a zeroth-order space within the initiator approximation. In practice, however, it gives essentially identical results to $\Ep$ with a smaller noise.

It is also interesting to consider the convergence of each estimator in a simulation. An example is shown in Fig.~(\ref{fig:formamide_converge}) for formamide in a cc-pVDZ basis and with a frozen core $(18\textrm{e}, 54\textrm{o})$. Preconditioning was used with parameters $\Delta \tau = 0.03$ au and $\Nspawn = 60$. The walker population was initialized from $10^6$ and grew to a final value of $6.3 \times 10^7$. It is found that $\Eref$ converges more slowly than $\Evar$. Note also that $\EvarPT$ is equal to $\Evar$ at initialization. This is because this definition of the PT2 correction only has contributions from spawnings cancelled due to the initiator criterion. All walkers are initialized within the deterministic space and therefore are initiators, and so the PT2 correction as defined in Eq.~(\ref{eq:en2_original}) is initially $0$. Meanwhile $\Ep$ initializes from a much lower energy since it takes the form $\bra \Phi | \hat{H} | \Psi \ket / \bra \Phi | \Psi \ket$, where $| \Phi \ket$ is immediately a much better estimate than $ | \Psi \ket$. However, both $\Ep$ and $\EvarPT$ converge to the same value once the simulation has equilibrated.

\subsection{Statistical error on perturbative corrections}
\label{sec:error_reduction}

Although $\EvarPT$ and $\Ep$ typically have a much smaller systematic (initiator) error than $\Eref$ and $\Evar$, they tend to have a much larger statistical error (noise). This is sometimes manageable, but becomes severe for large systems and small walker populations. To see why, consider the PT2 correction as it appears in $\EvarPT$:
\begin{equation}
\Delta E_2 = \frac{1}{(\Delta\tau)^2} \sum_a \frac{ S^1_a S^2_a }{ E - H_{aa} }.
\end{equation}
The summation is over all spawnings cancelled due to the initiator criteria. A similar term appears in estimators $\Ep$ and $\sigma^2$ (where the summation is performed over all spawnings, which does not affect the following argument).

The space sampled by the spawnings $S^1_a$ and $S^2_a$ contains up to double excitations from the occupied space, which is very large in general. Because replica sampling is required, a contribution to $\Delta E_2$ can only be made if spawnings from both replicas occur to the same determinant in the same iteration. As the space sampled becomes larger, or the number of spawned walkers becomes smaller, this becomes increasingly rare.

The preconditioned approach here allows one to perform fewer iterations ($\Niter$) with a larger number of spawning attempts per walker ($\Nspawn$). It can be shown that this approach leads to smaller noise on $\EvarPT$, $\Ep$ and $\sigma^2$, and improved efficiency overall. This can be seen by the following argument. Roughly, we expect the statistical error on an estimator such as $\Delta E_2$ to obey
\begin{equation}
\sigma_{\Delta E_2} \appropto \frac{1}{\sqrt{N_{\textrm{contribs}}}},
\end{equation}
where $N_{\textrm{contribs}}$ is the number of contributions to an estimate. Since a contribution is made only if two spawnings occur to the same determinant from two independent replicas, the number of contributions is roughly proportional to the density of spawnings,
\begin{equation}
N_{\textrm{contribs}} \appropto (\Nspawn)^2.
\end{equation}
This is an upper limit which will become less accurate as the space spawned to becomes saturated, i.e. for large $\Nspawn$ or a small number of orbitals. Assuming this holds, then
\begin{equation}
\sigma_{\Delta E_2} \appropto \frac{1}{\Nspawn}.
\end{equation}
However, for a total real simulation time $T$ the number of iterations performed scales as $\Niter \appropto T / \Nspawn $. Since $\Delta E_2$ is averaged over all iterations, we also have $\sigma_{\Delta E_2} \propto 1/\sqrt{\Niter}$. So for a constant simulation time $T$, as $\Nspawn$ is increased,
\begin{equation}
\sigma_{\Delta E_2} \appropto \frac{1}{\sqrt{\Nspawn}}
\label{eq:error_scaling}
\end{equation}
and the efficiency (with respect to estimation of $\Delta E_2$) follows
\begin{equation}
\epsilon_{\Delta E_2} = \frac{1}{\sigma^2 \times T} \appropto \Nspawn.
\end{equation}
Therefore, performing multiple spawning attempts per walker provides one way to greatly reduce the error on $\EvarPT$, $\Ep$ and $\sigma^2$. It should be emphasized that the following argument holds in FCIQMC both with and without preconditioning. However, preconditioning allows $\Delta \tau$ to be increased such that using a large value of $\Nspawn$ will not lead to slow convergence or a long autocorrelation time, which is critical. Therefore, preconditioning with large values of $\Nspawn$ and $\Delta \tau$ leads to a far more efficient algorithm overall.

\begin{figure}[t]
\includegraphics{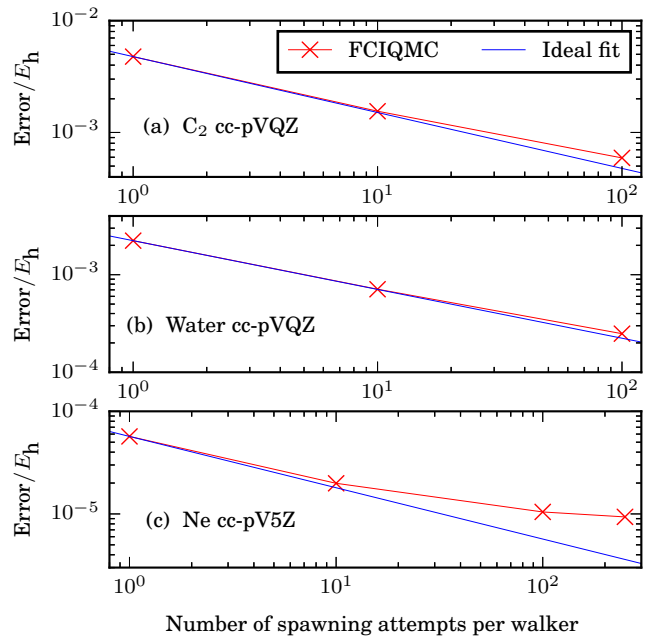}
\caption{Scaling of the statistical error estimate on $\Ep$ as $\Nspawn$ is increased, while keeping $\Nspawn \times \Niter$ fixed. (a) C$_2$ in a cc-pVQZ basis set at equilibrium bond length. (b) Water in a cc-pVQZ basis set. (c) Ne in a cc-pV5Z basis set. The frozen core approximation is used in each case. Ideal fits use the scaling motivated in the main text, relative to the value at $\Nspawn=1$.}
\label{fig:error_scaling}
\end{figure}

Fig.~(\ref{fig:error_scaling}) demonstrates the scaling of the statistical error estimate on $\Ep$ for three systems: C$_2$, cc-pVQZ at equilibrium bond length; Water, cc-pVQZ at equilibrium geometry; Ne, cc-pV5Z. Core electrons are frozen for each system. In each case $\Nspawn$ is increased while holding $\Nspawn \times \Niter$ constant and also holding $\Delta \tau \times \Niter$ constant (so that the final value of $\tau$ is fixed, and the total simulation time is approximately fixed). The reference population is held fixed as $\Nspawn$ is increased, leading to final walker populations that are very similar. It is seen that increasing $\Nspawn$ does indeed reduce the noise on $\Ep$. For example, with $\Nspawn=1$ the error estimate for C$_2$ is almost $5$ $\mEh$, which is reduced to $0.6$ $\mEh$ with $\Nspawn=100$, and the scaling of Eq.~(\ref{eq:error_scaling}) is approximately followed. This scaling is less accurate for Ne, where the number of orbitals is smaller (and so the space spawned to is smaller) and becomes saturated with spawned walkers more quickly.

\begin{figure*}[t]
\includegraphics{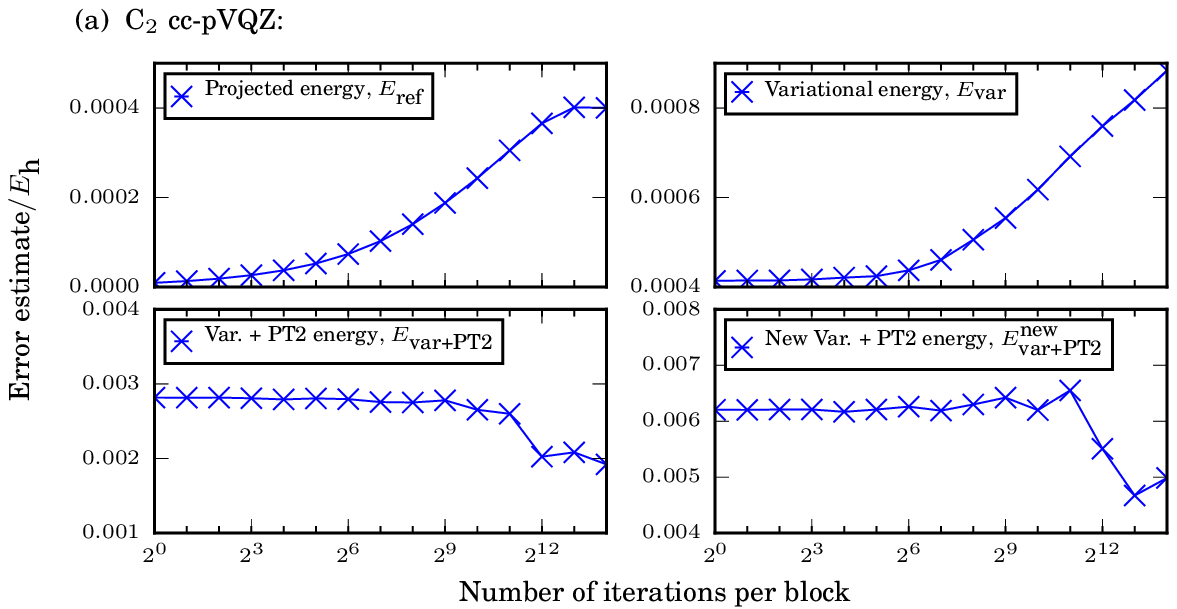}
\includegraphics{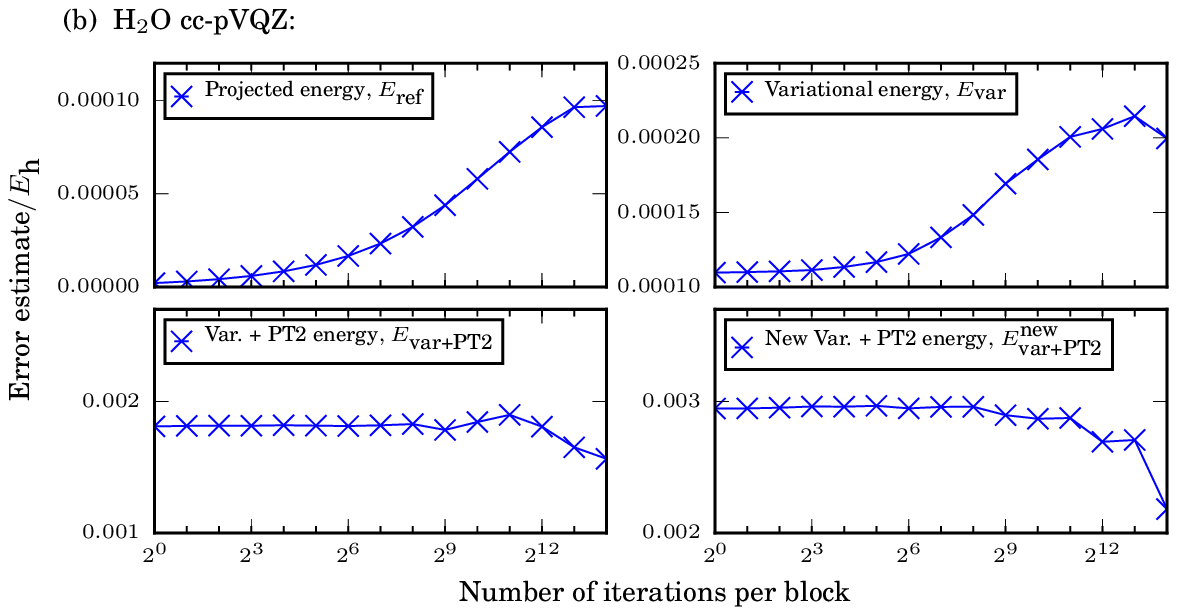}
\caption{Scaling of the statistical error estimate on the projected energy ($\Eref$), variational energy ($\Evar$), perturbatively corrected energy ($\EvarPT$), and preconditioned energy ($\Ep$) estimators. Errors are estimated by a blocking procedure (see the main text). For $\Eref$, the error is significantly underestimated with an uncorrelated analysis (a block length of $1$). For $\Evar$ this effect is lessened, but the error is nonetheless underestimated by a factor of $\sim 2$. In contrast, an accurate error estimate for perturbative quantities is obtained even with a block length of $1$ iteration. (a) C$_2$ at equilibrium bond length in a cc-pVQZ basis set. (b) Water in a cc-pVQZ basis set. The frozen core approximation is used for both systems.}
\label{fig:corr_length}
\end{figure*}

\subsection{Autocorrelation length on estimators}
\label{sec:corr_length}

Although $\EvarPT$ and $\Ep$ have larger noise than $\Eref$ and $\Evar$, they have a significant advantage regarding the correlation of QMC data. This is demonstrated in Fig.~(\ref{fig:corr_length}) where the two systems studied are C$_2$ and water, as defined in Section~\ref{sec:error_reduction}. To investigate the correlation of each estimator, we average each simulation into blocks of increasing length, and perform an uncorrelated error analysis using these blocks. This is simply the reblocking procedure, as described by Flyvbjerg and Petersen\cite{Flyvbjerg1989}. Note that for the simulations of C$_2$ and water, we took a total of $2^{18}$ and $2^{19}$ iterations to average over, respectively. Therefore even with a block length of $2^{14}$, we used $2^{4}$ or $2^{5}$ data points to construct error estimates, to ensure that these estimates are reliable.

If the data is correlated then the error estimate grows with increasing block length, eventually plateauing when subsequent blocks become approximately uncorrelated. This effect is seen to be most significant for $\Eref$, where for water performing an uncorrelated analysis gives an error estimate of $2.1 \times 10^{-6}$ $\Eh$, compared to a more realistic estimate of $1.2 \times 10^{-4}$ $\Eh$. An uncorrelated analysis of $\Evar$ gives an error estimate of $1.1 \times 10^{-4}$ $\Eh$ compared to an accurate estimate of $2.0 \times 10^{-4}$ $\Eh$, a much smaller but still non-negligible difference. We observe similar behavior across all systems investigated: an uncorrelated analysis typically underestimates the statistical error on $\Evar$ by a factor of $\sim 2$, while for $\Eref$ this factor is typically much larger.

For $\EvarPT$ and $\Ep$, the error estimate remains roughly constant as the block length is increased, indicating that data is approximately uncorrelated. We observe this across all systems studied. This is helpful, as a reliable error estimate on $\EvarPT$ and $\Ep$ may be obtained after a relatively small number of converged iterations. We suspect the reason for this is that the error on $\EvarPT$ and $\Ep$ is dominated by the term such as that in Eq.~(\ref{eq:en2_original}), involving a weighted dot product across the two spawning arrays. Although the FCIQMC wave function is heavily correlated from iteration to iteration, spawned walkers are essentially uncorrelated from each other. They are only correlated through their underlying dependence on the FCIQMC wave function, which should approximately cancel out in the denominator of the estimator. This seems to be very accurate based on our observations across many systems, although we would expect this observation to be only approximate theoretically. For example, $\EvarPT$ is formed as the sum of $\Evar$ and the PT2 correction; clearly an uncorrelated analysis is not exact for the former estimate (though $\Evar$ typically has a much smaller error than the PT2 term, perhaps explaining why this is not noticeable). We would therefore still recommend a protocol of performing many FCIQMC iterations when possible, but the situation is dramatically improved compared to that for $\Eref$.

Note that the above arguments do not depend using a large time step or preconditioning. A time step of $\Delta \tau = 10^{-3}$ au was used for both examples in Fig.~(\ref{fig:corr_length}). This small autocorrelation length on $\EvarPT$ and $\Ep$ is a property of the estimators themselves, and not the use of preconditioning.

\subsection{Convergence time in the preconditioned approach}
\label{sec:convergence}

As demonstrated in Figs.~(\ref{fig:converge_comp}) and (\ref{fig:variance}), the use of preconditioning allows a large time step to be used in FCIQMC. Typically one can set $\Delta \tau = 0.5$ au and achieve convergence without issue, which usually allows convergence within $20$ - $30$ iterations in our experience. Meanwhile, the original algorithm usually requires at least several thousand iterations to converge, and sometimes many more.

However, as discussed in Section~\ref{sec:precond_init}, setting a large time step also requires setting $\Nspawn$ to be very large. The simulation time in FCIQMC is dominated by the generation and processing of spawned walkers, such that iteration time is roughly proportional to $\Nspawn$. So for a fair comparison, we should instead look at convergence speed as a function of $\Niter \times \Nspawn$, rather than $\Niter$.

\begin{figure}[t]
\includegraphics{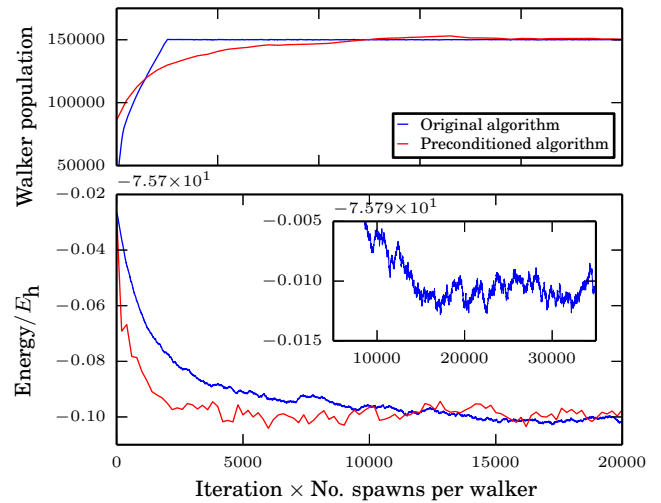}
\caption{An example comparison of convergence with and without preconditioning in FCIQMC, for C$_2$ in a cc-pVQZ basis, with a frozen core and at equilibrium bond length. Because the cost of each iteration is roughly proportional to the number of spawning attempts per walker ($\Nspawn$), convergence is plotted against the iteration number multiplied by $\Nspawn$. $\Nspawn=200$ with preconditioning, and $\Nspawn = 1$ without. The time step is set and updated to avoid bloom events, as implemented in NECI; the final values of $\Delta \tau$ with and without preconditioning are $0.34$ and $6.4 \times 10^{-4}$ au, respectively.}
\label{fig:c2_eq_converge}
\end{figure}

Another requirement for a fair comparison is that the time step should be chosen in a consistent manner. For this, we use the automatic system for choosing $\Delta \tau$, implemented in NECI. As discussed already, it is important that there are few bloom events, defined as an event where a spawned walker is created with weight greater than the initiator threshold ($n_a$). Allowing a large number of bloom events can greatly increase statistical noise and lower the efficiency of the algorithm. The automatic system in NECI looks for the largest bloom event from the previous iteration (if any), and reduces $\Delta \tau$ so that this spawning will have weight less than $n_a$ in future occurrences. The time step reaches a final value during convergence.

In Fig.~(\ref{fig:c2_eq_converge}), convergence is considered for the same system as in Fig.~(\ref{fig:converge_comp}) (C$_2$, cc-pVQZ, equilibrium geometry), but plotted against $\Niter \times \Nspawn$. With preconditioning we take parameters $\Nspawn=200$ and an initial time step of $\Delta \tau = 0.5$ au. For the original algorithm, we take $\Nspawn=1$ and an initial time step of $\Delta \tau = 0.0025$ au (so that the initial value of $\Delta \tau / \Nspawn$ is consistent). It can be seen that the benefit of preconditioning is now rather limited. In this case, the use of preconditioning speeds up convergence by only a small factor. We have tested this across a range of systems (including various basis set sizes, and both equilibrium and stretched regimes), and find that convergence is typically very similar between the two algorithms, by this metric.

It is important to understand why this is. Clearly, preconditioning is well established as improving the convergence rate considerably. As a function of number of iterations, convergence \emph{is} greatly sped up in FCIQMC. However in this stochastic setting the cost of each iteration scales strongly with the step size. This is dictated by the need to avoid large bloom events, to prevent large noise. Therefore, it is important to investigate bloom events more carefully. The unsigned weight of a spawned walker in the original algorithm is proportional to $\frac{1}{\Pgen(j \leftarrow i)} | H_{ji} |$, where $\Pgen(j \leftarrow i)$ is the probability of choosing determinant $| D_j \ket$, given spawning from $| D_i \ket $. With preconditioning this becomes proportional to $ \frac{ 1 }{ \Pgen(j \leftarrow i) } \, | \frac{H_{ji}}{E - H_{jj}} | $. Therefore the choice of $\Pgen(j \leftarrow i)$ is critical in determining the number of bloom events. In the original algorithm, the best choice of $\Pgen(j \leftarrow i)$ (allowing the maximum $\Delta \tau$ without bloom events) is given by
\begin{equation}
\Pgen(j \leftarrow i) = \frac{ |H_{ji}| }{ \sum_k |H_{ki}| }.
\label{eq:heat_bath}
\end{equation}
It is far too expensive to achieve this distribution exactly, but several schemes have been proposed to achieve this approximately. These include the heat bath approach of Holmes \emph{et al.}\cite{Holmes2016_1}, and approaches based on the Cauchy-Schwarz inequality (suggested by Alavi and co-workers\cite{smart_unpublished} and investigated recently by Neufeld and Thom\cite{Neufeld2018}). For preconditioned FCIQMC, the optimal choice of $\Pgen(j \leftarrow i)$ will be
\begin{equation}
\Pgen(j \leftarrow i) = \frac{ | \frac{ H_{ji}  }{ E - H_{jj} } | }{ \sum_k | \frac{ H_{ki} }{ E - H_{kk} } | }.
\label{eq:precond_dist}
\end{equation}
Therefore, optimal preconditioning requires a very different excitation generator to the original approach. In this study we have used Cauchy-Schwarz-based excitation generators implemented in NECI, designed to approximately achieve Eq.~(\ref{eq:heat_bath}), and so the above comparison gives a significant advantage to the original scheme. To see the problem, consider the simple example of water in a cc-pVDZ basis set with a frozen core. In this case, the correlation energy is $-0.215$ $\Eh$, and so any walker spawned to the HF determinant is amplified by a factor of $\sim 5$ by the preconditioner. Meanwhile, the largest value of $|E - H_{jj}|$ from a test simulation was $\sim 23$. Therefore the ratio of largest to smallest value of $|E - H_{jj}|$ is $\sim 100$ , and ideally we would like to make spawning to low-energy determinants $\sim 10$ times more likely, and spawning to high-energy determinants $\sim 10$ times less likely, relative to the current scheme. Doing so would allow the time step to be larger, and therefore convergence and autocorrelation times shorter, by this same factor (which will be system dependent). Alternatively, one could keep the time step fixed and reduce $\Nspawn$ by this factor, which would be particularly useful when the correlation length is short, or $\Delta \tau$ close to $1$ already.

Therefore, there is substantial potential to speed up the FCIQMC algorithm by modifying excitation generators for the preconditioned case. The design and optimization of excitation generators is an extensive task, which we do not consider here. Nonetheless, there is clearly a benefit to be gained in future work through this approach.

Lastly, in the above analysis we assumed that the iteration time scaled proportionally to $\Nspawn$. Actually, a large value of $\Nspawn$ is often more efficient than this. This is because some parts of the algorithm (such as the death step and deterministic projection) are independent of $\Nspawn$. For example, for benzene as studied in Section~\ref{sec:benzene}, the average iteration time divided by $\Nspawn$ is equal to 0.56 seconds without preconditioning and $\Nspawn=1$, while with preconditioning and $\Nspawn=150$ this value is equal to 0.41 seconds (with $\Nw = 1.28 \times 10^7$ in both cases). There are further ways in which large-$\Nspawn$ FCIQMC can be made more efficient, as discussed in the conclusion.

\begin{figure}[t]
\includegraphics{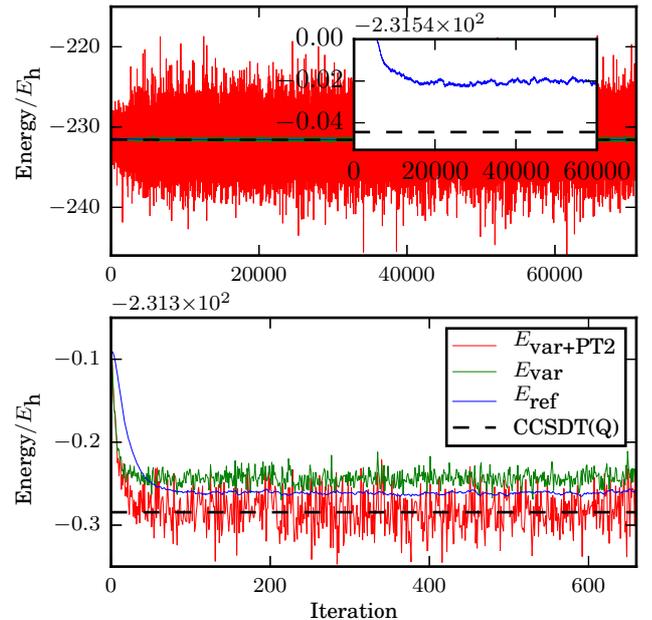}
\caption{Results for benzene in a cc-pVDZ basis with a frozen core $(30\textrm{e},108\textrm{o})$. Top: FCIQMC without preconditioning, with a time step of $1.9 \times 10^{-4}$ au and performing $1$ spawning attempt per walker. The inset shows the projected energy ($\Eref$) alone, with the dashed line showing the CCSDT(Q) result. Bottom: FCIQMC with preconditioning, with a time step of $0.1$ au and performing $150$ spawning attempts per walker. The walker population was $\Nw = 1.28 \times 10^7$ in both cases.}
\label{fig:benzene}
\end{figure}

\begin{table}[t]
\begin{center}
{\footnotesize
\begin{tabular}{@{\extracolsep{4pt}}lcc@{}}
\hline
\hline
Method & Estimator & Energy / $\Eh$ \\
\hline
CCSD(T)              &           & -0.5813 \\
CCSDT                &           & -0.5817 \\
CCSDT[Q]             &           & -0.5826 \\
CCSDT(Q)             &           & -0.5845 \\
\hline
FCIQMC               & $\Eref$   & -0.5609(3) \\
\: ($\Nspawn = 1$)   & $\Evar$   & -0.5420(5) \\
                     & $\EvarPT$ & -0.597(14) \\
\hline
FCIQMC               & $\Eref$   & -0.5612(3) \\
\: (preconditioned,  & $\Evar$   & -0.5435(5) \\
\: $\Nspawn = 150$)  & $\EvarPT$ & -0.5833(10) \\
\hline
\hline
\end{tabular}
}
\caption{Energies (shifted by $+231$ $\Eh$) for benzene in a cc-pVDZ basis set with a frozen core $(30\textrm{e},108\textrm{o})$. FCIQMC was performed without preconditioning ($\Nspawn=1$, $\Delta \tau = 1.9 \times 10^{-4}$ au), and with preconditioning ($\Nspawn=150$, $\Delta \tau = 0.1$ au). The FCIQMC simulations are those plotted in Fig.~(\ref{fig:benzene}). Relative to CCSDT(Q), $\Eref$ is too high by $\sim 23$ $\mEh$ and $\Evar$ by $\sim 41$ $\mEh$. For $\EvarPT$, the noise with $\Nspawn=1$ is too large for the estimate to be useful. With $\Nspawn=150$, a result similar to those from CCSDT[Q] and CCSDT(Q) is obtained. Note that initiator error in $\EvarPT$ is not fully removed here.}
\label{tab:benzene}
\end{center}
\end{table}

\subsection{Benzene}
\label{sec:benzene}

As an application of this approach to a larger system, we consider benzene in a cc-pVDZ basis set with a frozen core $(30\textrm{e},108\textrm{o})$, using the geometry of Ref.~(\onlinecite{Schreiber2008}). This is an example that would have been too challenging to study accurately with FCIQMC previously, even with significant computational resources, and so provides a good test.

Without preconditioning, parameters $\Delta \tau = 1.9 \times 10^{-4}$ au and $\Nspawn=1$ are chosen. With preconditioning, we take $\Delta \tau = 0.1$ au and $\Nspawn=150$. Both simulations used $1.28 \times 10^7$ walkers and were run for $11$ hours on $10$ $32$-core nodes, with $384$GB of RAM per node. These resources are modest compared to large-scale FCIQMC, which can be scaled up to more than $10^9$ walkers and $\sim 10^4$ CPU cores with appropriate load balancing\cite{HANDE}. Fig.~(\ref{fig:benzene}) presents the convergence of $\Eref$, $\Evar$ and $\EvarPT$ in both approaches. Table~\ref{tab:benzene} presents final estimates, averaged from convergence onwards, and compared to high-order coupled cluster.

Initiator error (relative to CCSDT(Q)) on $\Eref$ and $\Evar$ is roughly unchanged by the use of preconditioning. The estimate from $\Eref$ is too high by $\sim 23$ $\mEh$, while the estimate from $\Evar$ is too high by $\sim 41$ $\mEh$. With $\Nspawn=1$, the noise on $\EvarPT$ is too large to be useful. This is clear from Fig.~(\ref{fig:benzene}), where fluctuations from iteration to iteration are of size $\sim$ $20\Eh$. The final averaged value in Table~\ref{tab:benzene} has a stochastic error of $14$ $\mEh$, and no reliable conclusion can be made. Using $\Nspawn=150$, the noise is reduced substantially. Sensible convergence is seen, and the final estimate from $\EvarPT$ has an statistical error of $1$ $\mEh$. This energy is between the CCSDT[Q] and CCSDT(Q) results. Continuing the preconditioned simulation for a further $11$ hours increases the $\EvarPT$ estimate to $-231.5825(7)$ $\Eh$. Therefore, we suspect that the true $\EvarPT$ estimate (in the limit of zero statistical error) is slightly higher than that given in Table~\ref{tab:benzene}. The results here are not intended to be accurately converged FCI benchmarks, but estimates to assess the improvement made by $\EvarPT$ and the large-$\Nspawn$ approach. In this respect the approach described here makes a significant improvement over the previous method.

\section{Conclusion}
\label{sec:conclusion}

It has been demonstrated that FCIQMC can be performed with a preconditioner, in contrast to the traditional imaginary-time propagation, allowing time steps close to unity to be used. This results in a method which can typically converge within $20$ - $30$ iterations, while the original method typically requires at least several thousand iterations. In practice, the requirement that bloom events be avoided means that a large $\Nspawn$ must also be chosen. As a result, reductions in simulation time to convergence are rather more limited. This can be traced to the fact that currently-used excitation generators are optimized for imaginary-time propagation, and must be modified in the presence of a preconditioner. This will be an area for future work, and could greatly improve the speed of the method.

However, it has been shown that the use of a large $\Nspawn$ is a dramatic benefit for the calculations of perturbative corrections to initiator error. Such perturbative corrections improve the accuracy of the method dramatically, yet are almost free to calculate from rejected spawned walkers, so that we regard this as a clear improvement to FCIQMC. These improvements have been demonstrated for benzene $(30\textrm{e},108\textrm{o})$, which is certainly not feasible with the original i-FCIQMC algorithm, and where the PT2 correction was too noisy in the previous approach. Thus, while the preconditioned approach does not speed up convergence as one might expect, it is a significant benefit in the calculation of PT2 corrections to initiator error.

In practice, we also find that performing multiple spawning attempts per walker is more efficient in terms of iteration time per spawned walker. In future work, there are obvious ways in which the algorithm could be made more efficient in the large-$\Nspawn$ case. For example:
\begin{itemize}
\item In semi-stochastic FCIQMC, the deterministic projection is performed once per iteration. When performing a large number of cheap iterations (small $\Delta \tau$ and $\Nspawn$) this projection becomes too expensive beyond a deterministic space size of order $\sim 10^5$ or so. Using a small number of expensive iterations (large $\Delta \tau$ and $\Nspawn$), a much larger deterministic space could be used without this projection becoming the limiting cost.
\item The use of large $\Nspawn$ should make it more efficient to perform more of the algorithm deterministically, beyond the semi-stochastic approach. For example, for a determinant with $|C_i|$ walkers, it may be more efficient to generate all connected determinants and create spawnings accordingly, rather than calling the excitation generator $|C_i| \times \Nspawn$ times, particularly if this number is similar to or larger than the number of connected determinants (as is sometimes the case).
\end{itemize}
Combined with an excitation generator optimized for the preconditioned algorithm, such modifications could lead to a much faster algorithm. The use of perturbative estimators already allows a new range of systems to be studied accurately by the method. The implementation of these additional developments should allow the method to push further still in the near future.

\begin{acknowledgments}
N.S.B is grateful to St John's College, Cambridge for funding and supporting this work through a Research Fellowship. A.J.W.T. thanks the Royal Society for a University Research Fellowship under grant UF160398. C.J.C.S is grateful to the Sims Fund for a studentship. This study made use of the CSD3 Peta4 CPU cluster.
\end{acknowledgments}

\appendix

\section{A more rigorous PT2 correction to initiator error}
\label{sec:appendix}

As discussed in the text, a second-order perturbative correction to initiator error can be calculated by
\begin{equation}
\Delta E_2 = \frac{1}{(\Delta\tau)^2} \sum_a \frac{ S^1_a S^2_a }{ E_0 - H_{aa} },
\end{equation}
where the summation is performed over all spawnings which are cancelled due to the initiator criterion, and there is a normalization factor of $\bra \Psi^1 | \Psi^2 \ket$. This was motivated by selected CI methods. In such approaches, the Hamiltonian is diagonalized exactly in a variational subspace, allowing a perturbative correction to be calculated with an Epstein-Nesbet partitioning, and an analogous derivation was used\cite{Blunt2018} to obtain Eq.~(\ref{eq:en2_original}). However, the correction to i-FCIQMC is less rigorous because it is not simple to define a zeroth order space, and not clear that the FCIQMC wave function exactly samples the corresponding ground state.

To make the correction more rigorous, we present a second-order perturbative correction without considering a truncated space. This is the same approach used recently by Guo, Li and Chan\cite{Guo2018_2} and also by Sharma\cite{Sharma2018} to perturbatively correct a DMRG wave function, and we follow their idea.

Consider partitioning the Hamiltonian so that $\hat{H} = \hat{H}_0 + \hat{V}$ and $\hat{H}_0 |\Psi\ket = E_0 |\Psi\ket$, where $|\Psi\ket$ will be taken as the FCIQMC wave function (dropping the conventional $0$ subscript), and define $E_0 = \bra \Psi | \hat{H} | \Psi \ket$. The second-order energy correction can be obtained as
\begin{equation}
\Delta E_2 = - \bra \Psi | \hat{V} \hat{Q} [ \hat{H}_0 - E_0 ]^{-1} \hat{Q} \hat{V} | \Psi \ket,
\end{equation}
where $\hat{Q} = \mathbb{1} - | \Psi \ket \bra \Psi|$. An appropriate $\hat{H}_0$ can be defined by
\begin{equation}
\hat{H}_0 = \hat{P} E_0 \hat{P} + \hat{Q} \hat{H}_d \hat{Q},
\end{equation}
where $\hat{P} = | \Psi \ket \bra \Psi |$ and $\bra D_i | \hat{H}_d | D_j \ket = \delta_{ij} \bra D_i | \hat{H} | D_j \ket$ consists of the diagonal elements of $\hat{H}$ in the FCIQMC basis.

It can be shown that $\hat{Q} \hat{V} | \Psi \ket = (\hat{H} - E_0) | \Psi \ket$, and so
\begin{equation}
\Delta E_2 = - \bra \Psi | (\hat{H} - E_0) \; [ \hat{H}_0 - E_0 ]^{-1} (\hat{H} - E_0) | \Psi \ket.
\end{equation}
To proceed, one can make the approximation
\begin{equation}
\Delta E_2 = - \bra \Psi | (\hat{H} - E_0) \; [ \hat{H}_d - E_0 ]^{-1} (\hat{H} - E_0) | \Psi \ket,
\end{equation}
to avoid calculating the inverse of $[ \hat{H}_0 - E_0 ]$, which is not diagonal in the FCIQMC basis. This will be a very good approximation in this case, as $(\hat{H} - E_0) | \Psi \ket$ will be approximately zero in the space of occupied determinants. This inverse can be treated more carefully, as in Ref.~(\onlinecite{Guo2018_2}), but the above is more than sufficient for the FCIQMC case.

To put this in a form similar to that found for $\Ep$, we split the Hamiltonian into diagonal and off-diagonal components, $\hat{H} = \hat{H}_d + \hat{H}_{\textrm{off}}$. Then,
\begin{align}
\Delta E_2 &= - \bra \Psi | (\hat{H} - E_0) \; [ \hat{H}_d - E_0 ]^{-1} (\hat{H} - E_0) | \Psi \ket, \\
           &= - \bra \Psi | (\hat{H} - E_0) \; [ \hat{H}_d - E_0 ]^{-1} \hat{H}_{\textrm{off}} | \Psi \ket, \\
           &= - \bra \Psi | \; \hat{H}_{\textrm{off}} [ \hat{H}_d - E_0 ]^{-1} \hat{H}_{\textrm{off}} | \Psi \ket - \bra \Psi | \hat{H}_{\textrm{off}} | \Psi \ket.
\end{align}
where we used the definition of the zeroth-order energy, $\bra \Psi | (\hat{H} - E_0) | \Psi \ket = 0$. Finally, using $E_0 = \bra \Psi | \hat{H}_d + \hat{H}_{\textrm{off}} | \Psi \ket$,
\begin{equation}
E_0 + \Delta E_2 = - \bra \Psi | \hat{H}_{\textrm{off}} \; [ \hat{H}_d - E_0 ]^{-1} \hat{H}_{\textrm{off}} | \Psi \ket + \bra \Psi | \hat{H}_d | \Psi \ket.
\end{equation}
The spawned vector in FCIQMC can be written $S_i = - \Delta\tau \bra D_i | \hat{H}_{\textrm{off}} | \Psi \ket$, giving a final expression for the perturbatively corrected energy estimator in FCIQMC:
\begin{equation}
E_0 + \Delta E_2 = \frac{1}{(\Delta\tau)^2} \sum_i \frac{ S_i^1 S_i^2 }{ E - H_{ii} } + \sum_i C_i^1 H_{ii} C_i^2.
\end{equation}
We see that this expression includes the all terms in the perturbative correction of Eq.~(\ref{eq:en2_original}), and is almost identical to the expression for $\bra \Phi | \hat{H} | \Psi \ket$ obtained in Eq.~(\ref{eq:precond_energy}). The only difference is in the final term:
\begin{equation}
(E_0 + \Delta E_2) - \Ep = \frac{\sum_i C_i^1 H_{ii} C_i^2}{\sum_i C_i^1 C_i^2} - \frac{\sum_i \Phi_i^1 H_{ii} C_i^2}{\sum_i \Phi_i^1 C_i^2},
\end{equation}
where we have used the definition of $\Phi_i$ in Eq.~(\ref{eq:projected_wf_comp}), and included the required normalization factors. If $| \Psi \ket$ is approximately an eigenstate of $\hat{H}$, then $C_i \approx \Phi_i$, and the two estimators will give very similar results. Indeed, we find in practice that results from the two estimators are essentially identical after convergence.

\end{document}